\documentclass[%
reprint,
superscriptaddress,
amsmath, amssymb,
aps,
prb,
]{revtex4-2}

\usepackage[cp1252, utf8]{inputenc}
\usepackage{standalone}
\usepackage{graphicx}
\usepackage{dcolumn}
\usepackage{multirow}
\usepackage{diagbox}
\usepackage{bm}
\usepackage{braket}
\usepackage{dsfont}
\usepackage{mathtools}
\usepackage{color} 

\usepackage{svg}
\usepackage[%
  colorlinks=true,
  urlcolor=blue,
  linkcolor=blue,
  citecolor=blue
]{hyperref}
\usepackage{soul}
\usepackage{fixfoot}
\begin{document}

\title{Optical signatures of spin symmetries in unconventional magnets}
\date{\today}
\author{Javier Sivianes}
\affiliation{Centro de F\'{i}sica de Materiales (CSIC-UPV/EHU), 20018, Donostia-San Sebasti\'{a}n, Spain}
%
\author{Flaviano José dos Santos}
\affiliation{Theory and Simulation of Materials (THEOS), and National Centre for Computational Design and Discovery of Novel Materials (MARVEL), École Polytechnique Fédérale de Lausanne, 1015 Lausanne, Switzerland}
\affiliation{Laboratory for Materials Simulations (LMS), Paul Scherrer Institut, 5232 Villigen PSI, Switzerland}
\author{Julen Iba\~{n}ez-Azpiroz}%
\affiliation{Centro de F\'{i}sica de Materiales (CSIC-UPV/EHU), 20018, Donostia-San Sebasti\'{a}n, Spain}
\affiliation{IKERBASQUE, Basque Foundation for Science, 48009 Bilbao, Spain}
\affiliation{Donostia International Physics Center (DIPC), 20018
Donostia-San Sebasti\'{a}n, Spain}

\begin{abstract}

The concept of spin symmetries has gained renewed interest as a valuable tool for classifying unconventional magnetic phases, including altermagnets and recently identified  p-wave magnets.  In this work, we show that in compounds with weak spin-orbit coupling, the dominant
spin and charge
photoresponse is determined by spin group rather than the conventional magnetic group symmetry.
As a concrete realization we consider the nonlinear shift photocurrent in 
Mn$_5$Si$_3$, 
a material that features the two possible classes of
unconventional p-wave magnetism in the form of two competing 
spin structures, a coplanar and non-coplanar one.
While both are predicted to generate shift currents based on magnetic symmetry considerations, only the non-coplanar configuration 
survives the spin symmetry requirements. This is
numerically confirmed by 
our \textit{ab-initio} calculations, providing a protocol 
to experimentally identify the spin configuration 
of this promising material
in photogalvanic or transport measurements.

\end{abstract}

\maketitle

\emph{Introduction.}
The conventional understanding of magnetism has long been dominated by the paradigm of ferromagnetism and antiferromagnetism, characterized by the alignment or anti-alignment of magnetic moments, respectively. 
However, the emergence of new concepts such as 
altermagnetism~\cite{PhysRevX.12.031042, PhysRevX.12.040501} or 
unconventional p-wave magnetism~\cite{hellenes2024unconventional} challenges the traditional framework, introducing a renewed view on the subject.
Understanding and categorizing these 
novel phases  requires the notion of 
{spin group symmetries}~\cite{LITVIN1974538}; 
unlike magnetic group symmetries, they 
account for transformations that act independently on the spins and the lattice.
Although this separation is exact only in the non-relativistic limit, it serves as a useful approximation for systems with weak spin-orbit coupling
by introducing an additional degree of freedom~\cite{PhysRevX.12.021016}.

Altermagnets feature a compensated collinear magnetism that is distinguished from conventional antiferromagnetism by the appearance of a combined
symmetry involving time reversal ($\mathcal{T}$) and rotation~\cite{PhysRevX.12.031042, PhysRevX.12.040501}. 
$\mathcal{T}$ is therefore not a part of their point group, enabling $\mathcal{T}$-odd 
physics such as the anomalous Hall~\cite{Feng2022} and Nernst effects \cite{han2024observation, badura2024observation}
as well as optical excitations~\cite{adamantopoulos_spin_2024}. 
A  characteristic signature of altermagnets is the compensated spin-split Fermi surface displaying even-rotational symmetry~\cite{PhysRevX.12.031042, PhysRevX.12.040501,krempasky_altermagnetic_2024}.
In contrast, unconventional p-wave magnets are characterized by compensated noncollinear magnetism in parity($\mathcal{P}$)-odd systems where the combination of $\mathcal{T}$ and a fractional translation is part of the symmetry group~\cite{hellenes2024unconventional}. 
This leads to a $\mathcal{P}$-odd compensated Fermi surface with large  spin-splitting 
of $\sim 0.5$~eV
originating from non-relativistic sources. 

Notably, their inherent acentric structure make p-wave magnets
ideal platforms for studying the interplay between  noncollinear magnetism and $\mathcal{P}$-odd effects. Among the latter, 
nonlinear optical phenomena like the 
linear~\cite{fridkin_bulk_2001,sturman-book92,baltz-prb81} 
and circular photogalvanic effects~\cite{PhysRevB.61.5337}
have attracted great attention lately,
as they encompass fundamental physics such as quantized photoresponse in Weyl semimetals~\cite{dejuan2017,rees_helicity-dependent_2020,PhysRevB.102.121111} with 
potential applications in solar-energy harvesting~\cite{tan-cm16}. 
In particular, the shift current~\cite{PhysRevB.61.5337} is an intrinsic quadratic response that  
ranks amongst the largest contributions to
d.c. photocurrents~\cite{PhysRevB.107.L161403,PhysRevB.104.235203}, 
and exhibits close connection to geometric and topological
properties of materials~\cite{tan-cm16,morimoto-sa16}.


In this work, we study the role of spin symmetries in shaping the 
photoconductivity of unconventional magnets, extending beyond the 
standard magnetic point-group analysis. 
The framework applies broadly to responses at any order in the electric field, with the primary focus of the present paper on the quadratic shift current.
We establish the key principles in a minimal model, 
which we then apply to Mn$_5$Si$_3$, a candidate p-wave magnet with a debated noncollinear spin configuration below 60~K~\cite{10.1063/5.0156028, PhysRevB.105.104404}. 
By performing a detailed \textit{ab-initio} 
description of its 
optoelectronic  properties,
we demonstrate that the photoresponse is governed by 
an effective point group defined by the system's spin symmetry.
This provides a practical approach for determining the spin configuration 
of this intriguing material, and potentially other light-element magnetic compounds, through photogalvanic and transport measurements.


\emph{Minimal model.}
To illustrate how a noncollinear spin configuration and its symmetries 
 influence the quadratic shift photoconductivity $\sigma^{abc}$ \cite{sup}
 ($abc$ denote spatial indexes), we consider the  so-called double-exchange model~\cite{PhysRevB.91.054420, Kumar2005}:
%
\begin{align}
        H = &\sum_{< i,j >,\alpha}t_1 (c^{\dagger}_{i\alpha} c_{j\alpha} + h.c.) + \sum_{\ll i,j\gg,\alpha}t_2 (c^{\dagger}_{i\alpha} c_{j\alpha} + h.c.) \nonumber \\
        & +                       J\sum_{i, \alpha, \alpha^{\prime} }  \mathbf{m}^{i} \cdot c^{\dagger}_{i\alpha}\boldsymbol{S}_{\alpha\alpha^{\prime}}c_{i\alpha^{\prime}} \label{eq:model}
\end{align}
%
It consists of nearest (next-nearest) neighbor tunnelings $t_1$ ($t_2$)
and an on-site term describing the interaction between the 
atomic magnetic moment $\mathbf{m}^{i}$ 
and the electron spin  
through an exchange constant $J$, with $\mathbf{S}_{\alpha \alpha^{\prime}}$  the matrix elements of the Pauli matrices and 
$\alpha^{(\prime)}$ the spin projection onto a global axis. 

\begin{figure}
    \centering
    \includegraphics[width = 0.5\textwidth]{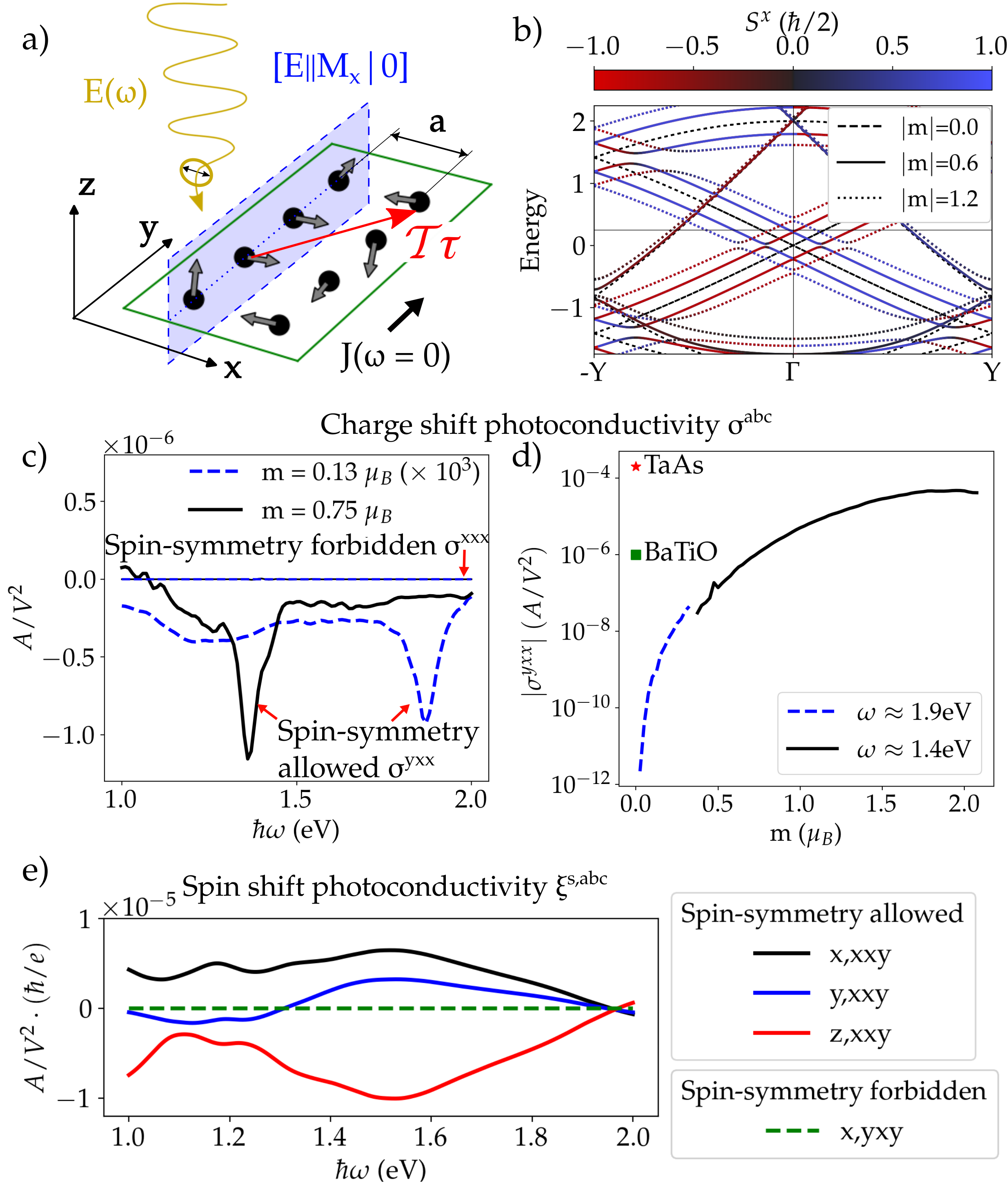}
    \caption{ a) Schematic illustration of 
    d.c. photocurrent generated on a rectangular lattice with noncollinear spin arrangement
    respecting $\mathcal{T}\boldsymbol{\tau}_m$ symmetry
    with $\boldsymbol{\tau}_m = (a,2a)$. 
    The spin symmetry $[E||M_{x}|\bm{0}]$ has been highlighted (see text).
    b) Spin polarized bands of the model for different values of $|\mathbf{m}|$ displaying $\mathcal{T}-$symmetry. c) Charge shift photoconductivity spectra for different values of $|\mathbf{m}|$.
    A $\times10^{3}$ factor has been applied to the $|\mathbf{m}|=0.13\mu_{B}$ curve for visibility purposes.
    d) Evolution of the value of  $\sigma^{yxx}$ for the peaks at $\omega \approx 1.4$ eV and $\omega \approx 1.9$ eV as a function of $|\mathbf{m}|$. 
    For reference, we have included maximum shift photoconductivity 
    of TaAs and BaTiO from Refs.~\cite{osterhoudt2019}
    and~\cite{PhysRevB.101.045104}.
    e) Spin shift photoconductivity spectra for $m=0.75\mu_{B}$, clearly fulfilling  the predictions of spin symmetries.}
    \label{fig:model}
\end{figure}

%
As a concrete example, we consider a two-dimensional square lattice with evenly spaced sites and a $\mathcal{P}$-breaking noncollinear magnetic structure as shown in Fig \ref{fig:model}a. 
We have picked typical parameter values for a 3d electron system: 
site-site distance $a=2$~Å,
tunneling coefficients
$t_1=1$~eV and $t_2=0.25$~eV, and 
exchange constant
$J=0.4$~eV/$\mu_B^2$ \cite{PhysRevB.16.255}.
The  lattice belongs to magnetic space group $P1.1'$, which only contains  the symmetry  $\mathcal{T}\boldsymbol{\tau}_m$, with $\boldsymbol{\tau}_m = (a, 2a)$ a fractional translation.  
The low level of symmetry implies that all components 
of the shift photoconductivity tensor, which transforms 
under point-group operations as the piezoelectric tensor~\cite{nye-book57}, 
are symmetry-allowed.

Consider now spin group symmetry operations~\cite{PhysRevX.12.021016,PhysRevX.14.031037}  denoted as $[R_s\|R_r|\boldsymbol{\tau}]$, where $R_s$ acts on spins alone and $[R_r|\boldsymbol{\tau}]$ is a conventional space group operation.
Since charge responses are
invariant under a pure spin operation $R_s$,
the shift tensor $\sigma^{abc}$ is constrained  by an \emph{effective point group} comprising only the spatial operations $R_r$.
In particular, the model of Fig.~\ref{fig:model}a 
is described by the spin symmetries $[\mathcal{T}||E|\bm{\tau}_m]$
and $[E||M_{x}|\bm{0}]$, which involve the 
real-space mirror operation $M_{x}$ 
and the spin-space identity $E$
[see Fig \ref{fig:model}a].
The spatial parts of these spin space symmetries conform the point group $m$, which restricts the 
independent components 
of the shift current to 
$\sigma^{yyy}$, $\sigma^{yxx}$ and $\sigma^{xxy}$~\cite{nye-book57}. 
Notably, our model calculations show that  
these components are several orders of magnitude 
larger than those that are forbidden by
spin-group symmetries, as illustrated 
for $\sigma^{yxx}$ (allowed) and
$\sigma^{xxx}$ (forbidden) 
in Fig.~\ref{fig:model}c
(see the rest of components in~\cite{sup}).
Note that this behavior  cannot be explained based on 
magnetic point group considerations, whereby all tensor components are allowed.

Let us next focus on 
the properties of the model 
as a function of the magnitude of the magnetic moment. 
In Fig. \ref{fig:model}b,  we show the corresponding  band structures. 
In the nonmagnetic limit of $|\mathbf{m}|=0$, the bands are Kramer-degenerate as a consequence of $\mathcal{PT}$ symmetry. The degeneracy 
is then lifted for $|\mathbf{m}|\neq 0$, producing 
non-collinearly spin-split bands that respect $E_{\mathbf{k}\uparrow} = E_{-\mathbf{k}\downarrow}$. 
The spin-splitting grows with increasing  $|\mathbf{m}|$, and in all cases the system is metallic.


The spectrum of the shift compoment
$\sigma^{yxx}$  is
illustrated in 
Fig. \ref{fig:model}c for  
$|\mathbf{m}|=0.13$ and $0.75~\mu_B$,  showing
clear peaks at photon frequencies $\omega_1\simeq$ 1.4~eV and $\omega_2\simeq1.9$~eV, respectively. 
We find that the peak at $\omega_1$
dominates for all $|\mathbf{m}|<0.33\,\mu_B$, while 
for $|\mathbf{m}|>0.38\,\mu_B$ the $\omega_2$ peak is the dominant one.
In Fig. \ref{fig:model}d, we have tracked the magnitude of these two peaks as a function of $|\mathbf{m}|$ in each of the dominance regions.
The figure unambiguously shows that the shift photoconductivity increases steadily with the size of the magnetic moment, while it vanishes in the  $|\mathbf{m}|\rightarrow 0$ limit. 
Only values larger than $|\mathbf{m}|\simeq1~\mu_B$ yield a substantial shift photoconductivity of the order of prototypical 
ferroelectrics like BaTiO~\cite{PhysRevB.101.045104}. 
For $|\mathbf{m}|\simeq1.5~\mu_B$
the photoconductivity reaches nearly $10^{-4}$~$\mathrm{A}/\mathrm{V}^{2}$, 
a notable value comparable to the highest reported to date in the 
Weyl semimetal TaAs~\cite{osterhoudt2019}.
Our results therefore indicate that noncollinear 
magnets not only support but can also generate 
substantial quadratic photocurrents, with the magnitude 
being highly sensitive to the size of the magnetic moments.

To conclude the model analysis, we briefly address spin photocurrents, focusing on the spin shift contribution, $\xi^{s,abc}$, where $s$ now
denotes spin index~\cite{PhysRevB.105.045201}. In the general case, determining the symmetry constraints on $\xi^{s,abc}$ requires considering transformations under both real-space and spin-space operations (see~\cite{sup} for formal details). 
However, the spin part of the symmetry $[E||M_{x}|\bm{0}]$ that characterizes the model in Eq.~\ref{eq:model} is the identity. Thus, 
as with the charge response, $\xi^{s,abc}$ is constrained by the 
mirror symmetry $M_x$ acting only in real-space degrees of freedom of $\xi^{s,abc}$,
which predicts that $\xi^{x,xxy}$, $\xi^{y,xxy}$, $\xi^{z,xxy}$ and $(b \leftrightarrow c)$ are the only spin-symmetry allowed tensor components. 
This differs from the result obtained through magnetic group symmetries, which permit all components.
As exemplified in Fig. \ref{fig:model}e (see rest of components in~\cite{sup}), the predictions based on spin symmetries are validated for the spin shift contribution as well, thereby expanding the scope and applicability of this approach to spin photocurrents.

\begin{figure}[!t] 
    
    
    \includegraphics[width=0.5\textwidth]{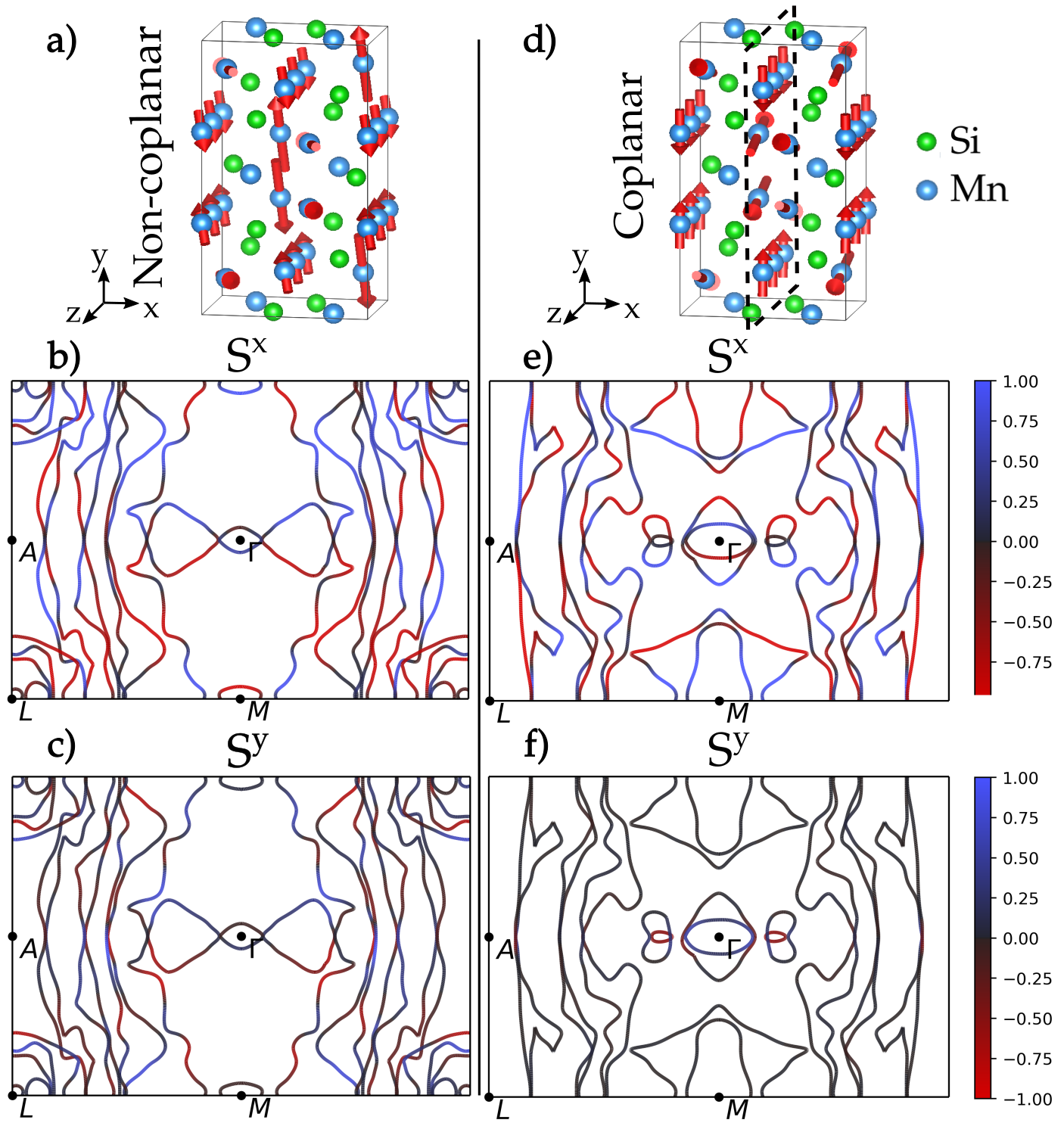}
    \caption{Left panel: Lattice (a) and Fermi surface at $k_y =0$ (b, c) of the non-coplanar structure for spin projections $S_{x}$ and $S_y$.  
    Right panel: same as left panel but for coplanar structure. In d), the plane of magnetization in $yz$ is displayed.  
    Spin projection $S_z$ not shown in neither panel; it is finite for the non-coplanar structure and virtually zero (like $S_y$) in the coplanar one. 
    \label{fig:material}}
\end{figure}

\emph{Mn$_5$Si$_3$: two candidate structures.}
The magnetic compound Mn$_5$Si$_3$  has received considerable 
attention lately thanks to its unconventional transport properties such as a large anomalous Hall effect \cite{10.1063/1.4943759} and topological Hall effect \cite{Sürgers2014}, or its inverse magnetocaloric effect \cite{PhysRevLett.120.257205}, making it an attractive platform for spintronics.
This material has zero net magnetization across all temperatures with three distinct magnetic phases: a paramagnetic phase for T $>$ 100~K~\cite{https://doi.org/10.1002/pssa.2210190145}; 
an antiferromagnetic collinear phase for 60 $<$ T $<$ 100~K, commonly referred to as AFM2 in the literature~\cite{https://doi.org/10.1002/pssb.2221580132, DEALMEIDA20092578, VINOKUROVA199596, PhysRevB.103.024407}; and a noncollinear phase  for T $<$ 60~K, 
AFM1~\cite{G_H_Lander_1967, P_J_Brown_1992, P_J_Brown_1995, M_Ramos_Silva_2002}.


Our focus is on the noncollinear phase, 
which falls into the category of p-wave magnets
cataloged recently~\cite{hellenes2024unconventional}.
The defining characteristic of this novel magnetic phase 
is the combination of $\mathcal{P}-$breaking noncollinear magnetic order and the symmetry $\mathcal{T}\boldsymbol{\tau}$, where  
$\boldsymbol{\tau}$ denotes a fractional translation. 
This results in spin-split bands with $E_{\mathbf{k}\uparrow} = E_{-\mathbf{k}\downarrow}$ for every polarization axis. 
There are two possible distinct classes of p-wave magnets. 
The first (class-I) is characterized by a 
coplanar magnetic configuration, where the 
magnetic moments lie on a given 
plane in real space.
As a result, in class-I  the combined symmetry [$C_{2\perp}\mathcal{T}\| E|\mathbf{0}]$  leaves the spin configuration invariant;  $C_{2\perp}$ describes a half rotation of the in-plane spin components at each site, while $\mathcal{T}$ flips the spins back to their original direction.
In turn, the class-II feature non-coplanar magnetic moments, which can lead to a large spin splitting induced by non-coplanar exchange contributions even in light-atom compounds~\cite{hellenes2024unconventional}.

The two classes of p-wave magnets come handy when characterizing 
the low-temperature structure of Mn$_5$Si$_3$, as
there is ongoing debate concerning its magnetic configuration. 
An early experimental work based in neutron polarimetry data~\cite{P_J_Brown_1992} suggested the arrangement shown in Fig. \ref{fig:material}a.
It belongs to the magnetic space group P$_S1= \{E, \mathcal{T}\tau_{xy}\}$ with $\boldsymbol{\tau}_{xy}=(1/2,1/2,0)$ and features non-coplanar magnetic moments, fitting into class-II.
More recently, the authors of Ref.~\cite{PhysRevB.105.104404} predicted an alternative spin configuration based on density
functional theory 
calculations followed by spin dynamic simulations. 
This structure is shown in Fig. \ref{fig:material}e and it is more symmetric, as it pertains to magnetic space group P$_C$222$_1$=  \{E, C$_{2y}$, C$_{2x}\boldsymbol{\tau}_z$, C$_{2y}\boldsymbol{\tau_z}$\} $\otimes$ $\{E, \mathcal{T}\boldsymbol{\tau}_{xy}$\} with $\boldsymbol{\tau}_{xy}=(1/2,1/2,0)$ and $\boldsymbol{\tau}_{z}=(0,0,1/2)$. In this case, all magnetic moments are located in the $yz$ plane and the $[C_{2x} \mathcal{T}|| E|\mathbf{0}]$ spin symmetry holds, fitting into class-I. 
The debate over the structure remains ongoing, with new 
experiments continually emerging \cite{sürgers2024anomalous}.


\emph{P-wave characteristics.}
We have computed the electronic properties of the two non-collinear structures of Mn$_5$Si$_3$
by means of
full-relativistic density functional theory calculations  as implemented in the Vienna ab initio package~\cite{PhysRevB.47.558, PhysRevB.54.11169, KRESSE199615} with PAW pseudopotentials \cite{PhysRevB.59.1758}. 
We have made use of the constrained local 
moments approach~\cite{PhysRevB.91.054420} 
to reach the desired magnetic configurations, 
yielding local magnetic moments of up to$~$2.6~$\mu_B$
that are in accordance with 
values estimated from experiment~\cite{P_J_Brown_1992} (see~\cite{sup}
for more details).


Fig. \ref{fig:material} shows the calculated band structure and the spin projected Fermi surface~\cite{Ganose2021} for the two configurations.
Both are metallic systems that show spin-split bands with inverse spin polarization $\boldsymbol{S}_{n}(\mathbf{k})=\bra{\psi_{\mathbf{k}n}} \boldsymbol{S} \ket{\psi_{\mathbf{k}n}}$ across the $\Gamma$ point as a consequence of the $\mathcal{T}\boldsymbol{\tau}$ symmetry.
However, while the non-coplanar structure displays finite spin polarization in all directions (see Fig. \ref{fig:material}c-d), 
the coplanar structure shows a negligible ${S}^{y}_{n}(\mathbf{k})$ and ${S}^{z}_{n}(\mathbf{k})$ components, except in the immediate vicinity of the $\Gamma$ point
(see Fig. \ref{fig:material}h for ${S}^{y}_{n}(\mathbf{k})$).
This implies that the k-space spin polarization is virtually 
perpendicular to the  plane of magnetization in real space,
a remarkable result that  can be
rationalized as follows~\cite{hellenes2024unconventional}. 
In the coplanar (class-I)  structure, 
the combined operation between 
$[C_{2x} T\| E|\mathbf{0}]$ and $T\boldsymbol{\tau}$
implies the existence of common eigenstates of  $  [C_{2x} \mathcal{T}\| E|\mathbf{0}]\circ [\mathcal{T}\|E|\boldsymbol{\tau}]$ and the 
crystal Hamiltonian. Then, noting that $\mathcal{T}^{2} = -1$ and $\boldsymbol{\tau} \boldsymbol{S} \boldsymbol{\tau^{-1}} = \boldsymbol{S}$, the parallel spin-polarization components are forced to
vanish given that 
\begin{equation}
\begin{split}\label{eq:spin_paral}
   {S}^{y,z}_{n}(\mathbf{k}) 
   = & \bra{\psi_{\mathbf{k}n}} C_{2x} S^{y,z} C_{2x}^{-1} \ket{\psi_{\mathbf{k}n}} = - {S}^{y,z}_{n}(\mathbf{k}),  
\end{split}
\end{equation}
whereas the $x$ component is unaffected by the spin rotation. 
This result offers a reliable approximation for materials with weak spin-orbit coupling, and
together with the electronic properties displayed in Fig.~\ref{fig:material}, 
showcases the unconventional p-wave nature of 
the two noncollinear spin-configurations of Mn$_5$Si$_3$.

\emph{Nonlinear optical fingerprint.}
We turn next to analyze the quadratic shift current 
of Mn$_5$Si$_3$.
According to the magnetic point group, the shift current 
is allowed in the two structures depicted in Fig.~\ref{fig:material};
the coplanar configuration enables three finite and independent components ($\sigma^{xyz},\sigma^{yzx}$ and $\sigma^{zxy}$), while for the non-coplanar configuration all 18 tensor components are allowed and independent.
In contrast, spin-group considerations severely restrict  the allowed 
photoconductivity components. 
The effective point group of the 
non-coplanar structure is $m$, which implies that $\sigma^{abc}$ should vanish 
for odd instances of the $z$ component. 
As for the coplanar structure, its effective point
group $mmm$ actually restores $\mathcal{P}$, 
which completely suppresses the non-relativistic contributions to quadratic photoresponses.



\begin{figure} 
     \centering
     \includegraphics[width = 0.5\textwidth]{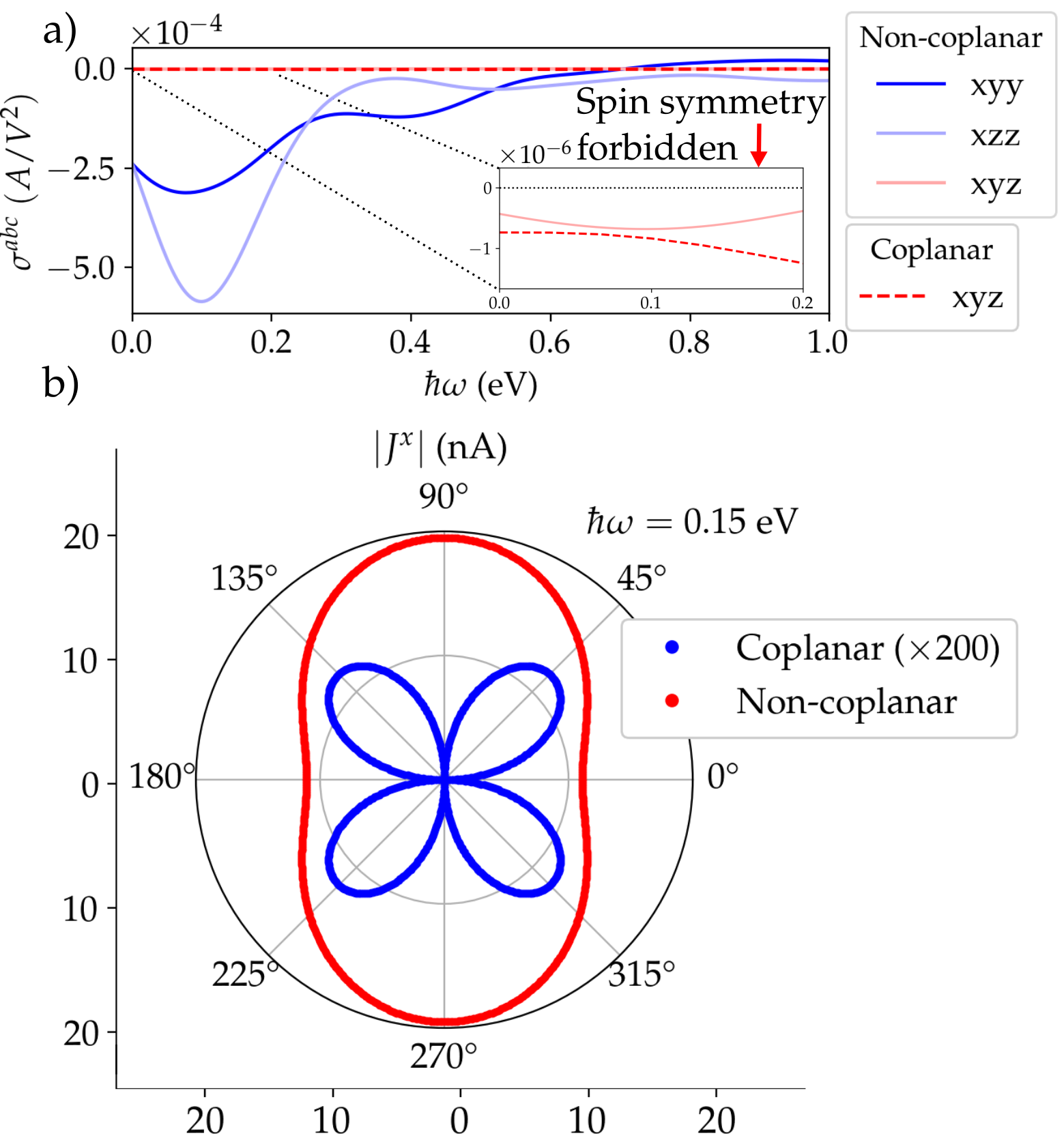}
     \caption{a) Top and bottom: shift photoconductivity of the non-coplanar 
     and coplanar structures of Mn$_5$Si$_3$, respectively. 
     b) 
     D.c. photocurrent $J^{x}$ at $\hbar\omega=0.15$~eV 
     as a function of polarization angle in the two structures.
     \label{fig:sc}}
\end{figure}

The numerical evaluation of the shift photoconductivity 
up to a photon frequency  of $1$~eV
is illustrated in Fig. \ref{fig:sc}a  
for the non-coplanar and coplanar structures.
For conciseness, we have focused on  photocurrent generated along 
$x$  under light polarized linearly in the $yz$ plane, hence
the involved tensor components are 
$\sigma^{xyy}$, $\sigma^{xzz}$ and $\sigma^{xyz}$
(see \cite{sup} for the complete set of components, 
as well as the linear response). 
The key component is $\sigma^{xyz}$, 
which  is allowed by the 
magnetic group but forbidden by the spin group
in both the non-coplanar and coplanar structures.

Fig. \ref{fig:sc}a shows that the predictions of spin-symmetry
arguments are largely fulfilled,
given that $\sigma^{xyz}$ of both configurations is 
two orders of magnitude smaller than the 
finite components of the non-coplanar structure. 
This characteristic cannot be captured by conventional magnetic point-group considerations, highlighting the effectiveness of spin symmetries in predicting the response properties of  magnets with weak spin-orbit coupling. 
As for the magnitude, 
the shift photoconductivity of the non-coplanar configuration
peaks at 
$\sim10^{-4}$ A/V$^{2}$, which
is an order of magnitude larger than 
semiconductors 
such as GaAs~\cite{PhysRevB.74.035201} or
BC$_2$N~\cite{PhysRevResearch.2.013263,10.21468/SciPostPhys.12.2.070}, and comparable 
to Weyl semimetals 
TaAs and TaIrTe$_4$, which rank amongst the largest reported to date in any material~\cite{osterhoudt2019,Ma2019,PhysRevB.97.241118,PhysRevB.107.205204}.




The strong difference between 
the shift-current magnitude of the co-planar and non-coplanar structures 
can be used as a probe to experimentally discern the actual spin configuration of Mn$_5$Si$_3$.
We propose the measurement of the d.c.
photocurrent for this purpose,
which is illustrated 
in  
Fig.~\ref{fig:sc}b
as a function of the 
incident polarization angle 
$\mathbf{E}(\omega)=E_0(\omega)(0 ,\cos(\theta), \sin(\theta))$
(see \cite{sup} for calculation details). 
%
%
%
We have chosen a photon frequency of 0.15 eV and an intensity of $480$~nW distributed over a spot size of $w=5~\mu$m, values that are well inside the experimentally 
accessible range~\cite{Ma2019,Zhang2019}. 

The figure shows that the non-coplanar structure generates a photocurrent of approximately 20 nA that is well within the measurable limits~\cite{Ma2019,Zhang2019}, while the coplanar structure produces a photocurrent two orders of magnitude smaller, consistent with the photoconductivity components shown in 
Fig.~\ref{fig:sc}a. 
The polar distribution differs significantly between the two cases: the non-coplanar structure follows a $\sigma^{xyy}\cos^{2}\theta + \sigma^{xzz}\sin^{2}\theta$ pattern, whereas the coplanar structure exhibits a $\sigma^{xyz}\cos\theta\sin\theta$ shape. This substantial difference in magnitude and angular form provides a clear method to identify the underlying spin configuration in photogalvanic 
measurements. Alternatively, other nonlinear effects, such as second harmonic generation~\cite{lee_magnetic_2021,PhysRevResearch.4.L022022,PhysRevB.104.L180407}, could also be used to determine the material’s symmetries based on 
similar principles.


\emph{Discussion}. In this work we have applied the concept of spin symmetries
to the  photoresponse of noncollinear magnets, showing that 
it imposes constraints beyond conventional magnetic group considerations.
This approach is relevant in materials with weak spin-orbit coupling, where spin and lattice degrees of freedom are effectively decoupled
and spin symmetries thereby provide an adequate 
description. 
Recently, over fifty thousand spin group types encompassing nearly two thousand compounds have been identified~\cite{PhysRevX.14.031037}, providing a broad field for exploring their optical properties.

As a case study, we have focused on 
the recently cataloged 
unconventional p-wave magnets and showed that they can 
produce a quadratic shift photocurrent that is symmetry-enabled by their 
spin configuration. 
Unlike most photovoltaic effects, the charge response scales with the magnitude of local magnetic moments,  reaching levels comparable to those in Weyl semimetals. 
This relationship opens the possibility to design materials with enhanced responses along specific directions, \textit{e.g.}, by considering magnetic 
configurations with predominant directionality of their spins.

We have examined Mn$_5$Si$_3$ as a concrete material realization,
which adopts a compensated noncollinear magnetic structure 
below 60~K  with two proposed spin configurations -- non-coplanar and coplanar -- each exhibiting distinct symmetry properties. 
While magnetic point-group considerations predict a finite shift current for both structures, spin symmetries confine this effect exclusively to the non-coplanar configuration.
The spin symmetry prediction is confirmed by our \textit{ab-initio} 
calculations and provides a tangible method for identifying the spin configuration of this promising material through optical 
and transport measurements~\cite{PhysRevB.92.165424}.

The results presented in this manuscript  are 
not limited to unconventional magnets but are expected
to hold relevance in any light-element magnet where spin symmetries
provide additional information beyond that contained in the magnetic point group.
Moreover, the influence of spin symmetries extends across responses at all orders of the electric field; this includes linear contributions 
such as the anomalous Hall
effect, which is a signature of altermagnets~\cite{Feng2022}. 
Besides affecting the dominant light-absorption coefficients, 
the consequences are likely to propagate into the behavior 
of collective excitations such as plasmons and magnons, thereby 
affecting properties like
reflectance and spin transport, \textit{e.g.}, spin Seebeck and spin Nernst effects. Finally, our findings
on the spin counterpart of the effect lay the groundwork for a comprehensive classification, which is expected to have implications in timely topics such
as the generation of pure bulk spin photocurrents~\cite{xu_pure_2021}.

\textit{Acknowledgements.} We thank useful feedback from F. de Juan, S. Tsirkin, N. Marzari and I. Errea. 
This project has received funding from the European Union’s Horizon 2020 research and innovation programme under the European Research Council (ERC) grant agreement No 946629 StG PhotoNow.

\bibliography{bibliography}
\setcounter{equation}{0}
\setcounter{figure}{0}
\renewcommand{\thesubsection}{S\arabic{subsection}}
\renewcommand{\thetable}{S\arabic{table}}
\renewcommand{\thefigure}{S\arabic{figure}}
\renewcommand{\theequation}{S\arabic{equation}}
\newpage
\onecolumngrid
\section*{Supplementary material} \label{sec: supplementary}

\subsection{Details on the calculation of the shift current}
\subsubsection{Charge shift current}
The (charge) shift current describes a d.c. photocurrent generated in acentric  materials.
In the length gauge, it is characterized by the photoconductivity tensor~\cite{PhysRevB.61.5337}
\begin{eqnarray}\label{eq:shift}
    \sigma^{abc}(\omega) = &&-\frac{i\pi{e}^3}{2\hbar^2}\int \frac{d\mathbf{k}}{(2\pi)^3}\sum_{mn}f_{nm}(I_{mn}^{abc} + I_{mn}^{acb})\left[\delta\left(\omega_{nm}-\omega\right)+\delta\left(\omega_{mn}-\omega\right)\right].
\end{eqnarray}
Here $a,b$ and $c$ are Cartesian components, $m$ and $n$ are band indexes, $f_{nm}=f_{n}-f_{m}$ is the occupation factor difference, $\omega_{mn}=\omega_{m}-\omega_{n}$ is the energy gap of the bands involved and the transition matrix element $I_{mn}^{abc}=r_{mn}^{b}r_{nm}^{c;a}$
contains the dipole term $r_{mn}^{b}=i(1-\delta_{nm})\bra{n}\partial_{a}\ket{m}$ and its generalized derivative $r_{nm}^{c;a}=\partial_a r_{nm}^{c}-i\left(A_{nn}^{a}-A_{mm}^{a}\right)r_{nm}^{c}$ with $A_{nn}^{a}=i\bra{n}\partial_{a}\ket{n}$ the intraband Berry connection. 
The shift current results of the double-exchange model shown
in the main text have been calculated~\cite{Tsirkin2021} by applying the Wannier 
interpolation technique to Eq. \ref{eq:shift}, as described in~\cite{PhysRevB.97.245143}.

For the shift-current calculations of Mn$_5$Si$_3$, we have 
found it more convenient to work directly with the Bloch eigenstates
rather than constructing  Wannier functions.
The reason is that the standard minimization procedure to obtain maximally  localized
Wannier functions 
does not enforce crystal symmetries. 
In materials with a non-trivial band structure where the 
minimization procedure is challenging, 
like in Mn$_5$Si$_3$, this 
procedure can produce an effect on
symmetry-forbidden tensor components whereby they 
become numerically finite~\cite{wang_first-principles_2019}. 
Given the importance of symmetry considerations in our work, 
we have chosen to proceed by considering the alternative velocity gauge expression for 
the shift photoconductivity~\cite{doi:10.1021/acs.jctc.3c00917} 
\begin{equation}
\sigma^{abc}(\omega) =  \frac{\pi |e|^{3}}{2 \hbar^{2} V} \sum_{\bm{k}\in \text{BZ}}\sum_{m,n} \frac{f_{m,n}}{\omega^{2}_{m,n}} \text{Im} \left[ v^{b}_{m,n}\sum_{l\neq m,n} \left( \frac{v^{a}_{n,l} v^{c}_{l,m}}{\omega_{n,l}} - \frac{v^{c}_{n,l}  v^{a}_{l,m}}{\omega_{lm}} \right) + (b \leftrightarrow c) \right] \delta (\omega_{m,n} - \omega),\label{eq: velocity}
\end{equation}
where we have calculated all quantities involved directly from Bloch eigenstates,
without employing Wannier functions.
Eq.~\ref{eq: velocity} avoids the troublesome k-space derivatives 
involved in the dipole term at the cost of introducing a truncation error in the band summations. 
We have employed a basis of 300 bands to converge  
the calculation of the shift current
in the two structures of Mn$_5$Si$_3$ analyzed in the main text. 
The considered k-mesh has been 19x15x21, which  is a modest number if compared to
usual values obtainable using Wannier interpolation~\cite{PhysRevB.97.245143}.
In return, this way of proceeding makes 
sure that the symmetry of the Bloch functions inherited 
from crystal symmetries is correctly translated  into 
the photoconductivity components.

\subsubsection{Spin shift current}
The spin shift current is the spin version of the shift current described in \cite{PhysRevB.105.045201} as
\begin{equation}
    \xi^{s,abc}_{shift}=\frac{i\pi q^{3}}{2\hbar^{2}V\Omega}\sum_{\mathbf{k},m,n}f_{mn} I^{s,abc}_{mn}(\mathbf{k}) \delta(\Omega+\omega_{mn})
\end{equation}

with  $I^{s,abc}_{mn}(\mathbf{k}) = d^{s,ba}_{mn}v^{c}_{nm} - d^{s,ca}_{nm}v^{b}_{mn}$ the associated transition matrix element,
$d^{s,ba} = j^{s,ab}_{mn} + \sum_{p\neq m} \frac{j^{s,a} v^{b}_{pn}}{\omega_{mp}} + \sum_{p \neq n} \frac{v^{b}_{mp}j^{s,a}_{pn}}{\omega_{np}}$,   $j^{s,ab} = \frac{1}{2}\bra{u_m} \{S^s, \partial^a \partial^b H \} \ket{u_n} $,
$j^{s,a} = \frac{1}{2}\bra{u_m}\{S^s, v^a\} \ket{u_n}$, 
$S^s$ the spin operator,  $f_{mn}=f_{m} - f_{n}$ the occupation factor difference, $\Omega$ the frequency of the incident light and $\omega_{mn}$ the energy difference of the bands. 

In particular we consider the spin shift current that arises under circularly polarized light obtained as 
\begin{equation}
        \xi^{s,abc}_{C} = \text{Im}( \xi^{s,abc} -  \xi^{s,acb})
\end{equation}
which is antisymmetric in $(b \leftrightarrow c)$ and $\mathcal{T}$ even.

\subsection{Spin symmetries in model calculation}

\begin{figure}[h!]
    \centering
    \includegraphics[width=1\linewidth]{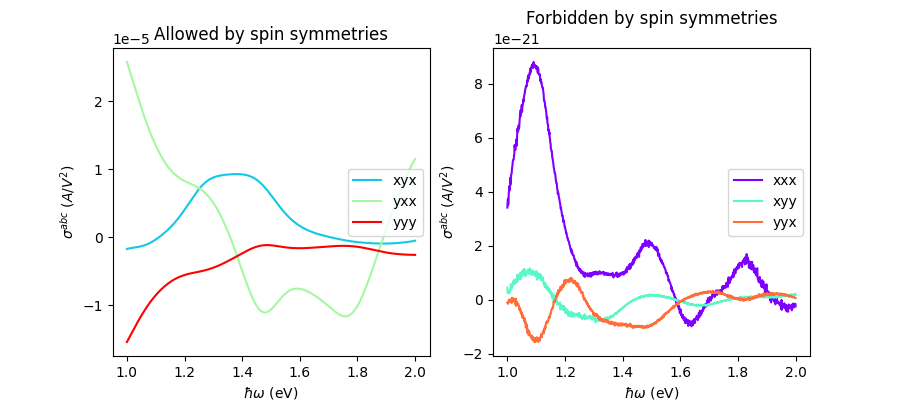}
    \caption{Charge shift photoconductivity spectrum of the model described in the main text for $m=2\mu_B$.}
    \label{fig:sup_model}
\end{figure}

\begin{figure}[h!]
    \centering
    \includegraphics[width=1\linewidth]{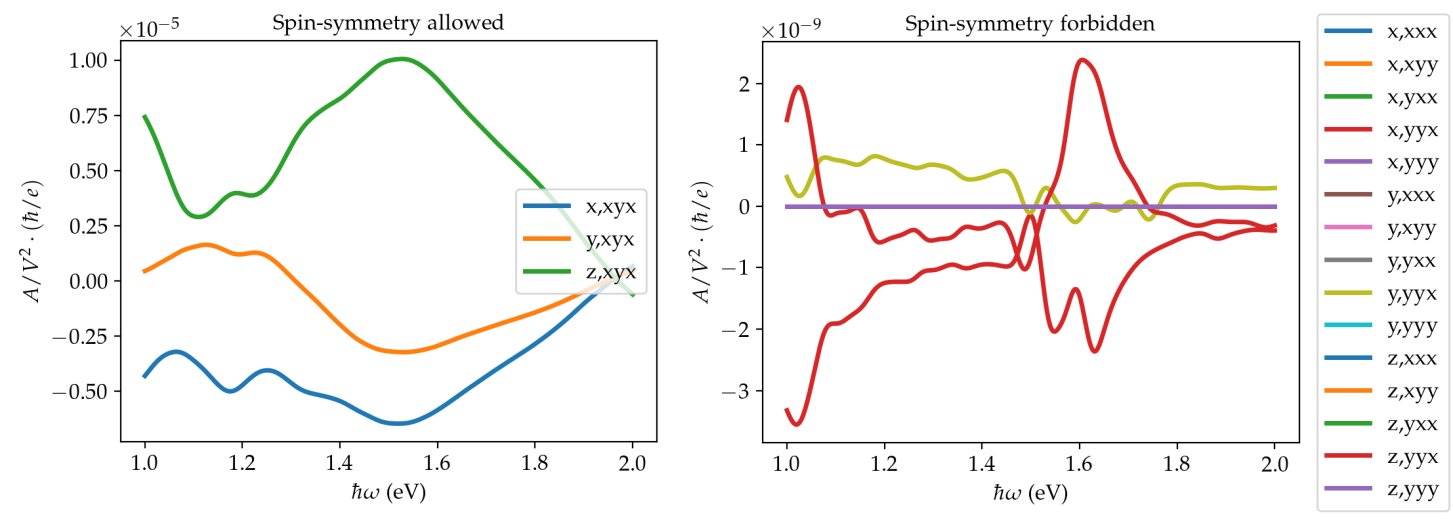}
    \caption{Spin shift photoconductivity spectrum of the model described in the main text for $m=0.75\mu_B$.}
    \label{fig:sup_model_spin}
\end{figure}

Fig. \ref{fig:sup_model} and  Fig. \ref{fig:sup_model_spin} show charge and spin the shift photoconductivity spectrum of the model described in the main text for $m=2\mu_B$ and $m=0.75\mu_B$ respectively, clearly displaying the impact of the effective point group $m$ arising from spin symmetry considerations, whereby the forbidden components are more than 4 orders of magnitude smaller than the allowed ones. We remind that magnetic group symmetries allow all components. 

\subsection{Structural and computational details for the calculations in Mn$_5$Si$_3$}

\paragraph{Non-coplanar structure}
Both atomic positions and magnetic moments can be found in the MAGNDATA database \cite{Gallego:ks5530, Gallego:ks5532} at Bilbao Crystalographic Server.
Bellow we present the atomic positions and the magnetic moments from the DFT calculations, where the first 20 positions correspond to the Mn atoms followed by the 12 Si atoms.

\begin{table}[h!] 
    \begin{minipage}[t]{0.45\linewidth}
        \centering
        \caption{Mn atoms}
        \begin{tabular}{|c|c|c|c|c|c|}
            \hline
         x&  y&  z&  $m_x$ ($\mu_B$)&  $m_y$ ($\mu_B$)& $m_z$ ($\mu_B$)\\ \hline
         0.00&  0.33&  0.00& -0.293& 1.088&-0.554\\ 
         0.50&  0.17&  0.50&-0.293& 1.086&-0.555\\ 
         0.50&  0.17&  0.00&-0.293& 1.086&-0.555\\ 
         0.00&  0.33&  0.50& -0.293& 1.088&-0.554\\ 
         0.50&  0.83&  0.00&  0.293&-1.086& 0.554\\ 
         0.00&  0.67&  0.50&  0.293&-1.087& 0.554\\ 
         0.00&  0.67&  0.00&  0.293&-1.087& 0.554\\ 
         0.50&  0.83&  0.50&  0.293&-1.086& 0.554\\ 
         0.24&  0.00&  0.25&  0.000& 0.000& 0.001\\ 
         0.74&  0.50&  0.25& -0.000&-0.000&-0.000\\ 
         0.76&  0.00&  0.75&  0.000& 0.000& 0.000\\ 
         0.26&  0.50&  0.75& -0.000&-0.024&-0.001\\ 
         0.88&  0.12&  0.25&  0.121&-2.477&-0.919\\ 
         0.38&  0.38&  0.25&  0.121&-2.476&-0.918\\ 
         0.38&  0.12&  0.25& -0.121& 2.477& 0.919\\ 
         0.88&  0.88&  0.25& -0.121& 2.475& 0.918\\ 
         0.12&  0.88&  0.75& -0.931&-0.198&-2.485\\ 
         0.62&  0.62&  0.75& -0.930&-0.199&-2.484\\ 
         0.62&  0.38&  0.75&  0.931& 0.199& 2.486\\ 
         0.12&  0.12&  0.75&  0.930& 0.198& 2.484\\ 
         \hline
        \end{tabular}\label{sup: Mn}
    \end{minipage}
    \hfill
    \begin{minipage}[t]{0.40\linewidth}
        \centering
        \caption{Si atoms}
        \begin{tabular}{|c|c|c|c|c|c|}
        \hline
         x&  y&  z&  $m_x$ ($\mu_B$)&  $m_y$ ($\mu_B$)& $m_z$ ($\mu_B$)\\ \hline
         0.60&  0.00&  0.25&  0.00&  0.00& 0.00\\ 
         0.10&  0.50&  0.25&  0.00&  0.00& 0.00\\ 
         0.40&  0.00&  0.75&  0.00&  0.00& 0.00\\ 
         0.90&  0.50&  0.75&  0.00&  0.00& 0.00\\ 
         0.70&  0.03&  0.25&  0.00&  0.00& 0.00\\ 
         0.20&  0.80&  0.25&  0.00&  0.00& 0.00\\ 
         0.70&  0.70&  0.25&  0.00&  0.00& 0.00\\ 
         0.20&  0.20&  0.25&  0.00&  0.00& 0.00\\ 
         0.30&  0.70&  0.75&  0.00&  0.00& 0.00\\ 
         0.80&  0.20&  0.75&  0.00&  0.00& 0.00\\ 
         0.30&  0.30&  0.75&  0.00&  0.00& 0.00\\ 
         0.80&  0.80&  0.75&  0.00&  0.00& 0.00\\ 
         \hline
        \end{tabular}\label{sup: Si}
    \end{minipage}
\end{table}

\begin{figure}
    \centering
    \includegraphics[width=1\linewidth]{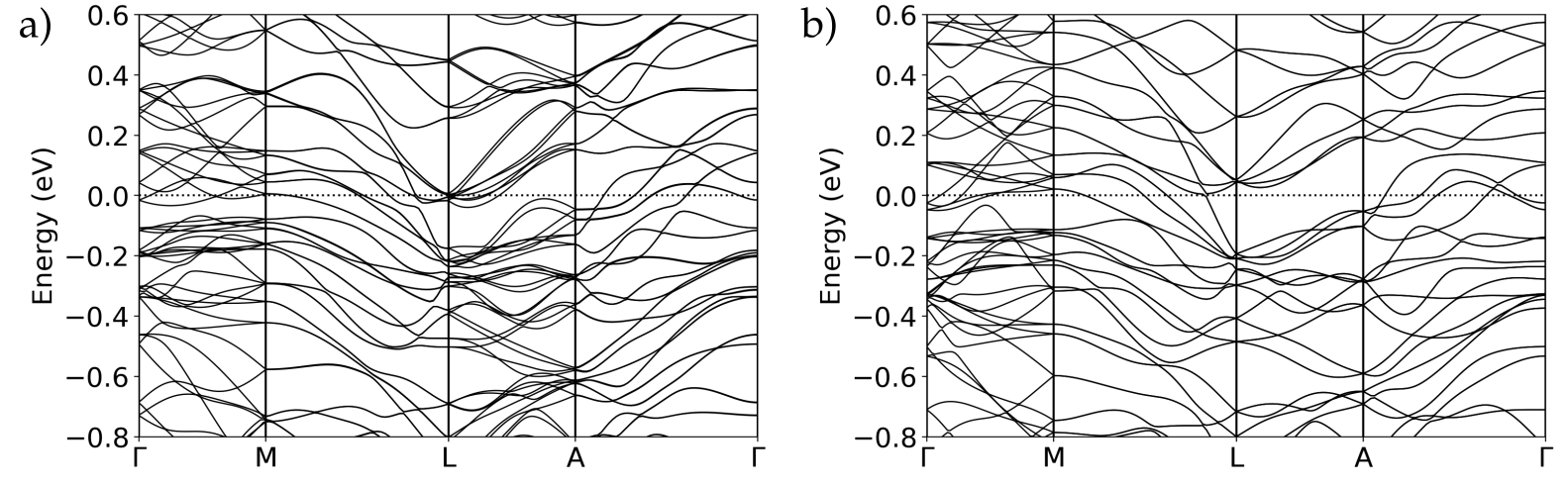}
    \caption{Band structure of the (a) non-coplanar and (b) coplanar structures of Mn$_5$Si$_3$. The horizontal dotted line marks the Fermi level.}
    \label{fig:bands}
\end{figure}

The DFT calculation were converged to $10^{-7}$ eV, for which we used a energy cutoff of 440eV, an smearing of 0.05eV and a gamma-centered kpoint mesh of $9\times5\times11$. In our calculations we set the initial magnetic moments to a value $\sim$25\% higher and let them converge to the self-consistent solution shown in Table \ref{sup: Mn}, \ref{sup: Si}. Additionally, the direction of the magnetic moments were constrained with a penalty weight of 10 and spheres of radius 1Å were considered for the on-site magnetization. The DFT result for the total energy is $E_T = -254.28$ eV and the band structure can be found in \ref{fig:bands}.




\paragraph{Coplanar structure}
\begin{table}[h!] 
    \centering
    \begin{tabular}{|c|c|c|c|c|c|}
    \hline
         x&  y&  z&  $m_x$ ($\mu_B$)& $m_y$ ($\mu_B$)&$m_z$ ($\mu_B$)\\ \hline
         0.00&  0.33&  0.00&  0.00& -1.378&0.00\\ 
         0.50&  0.17&  0.50&  0.00& -1.378&0.00\\ 
         0.50&  0.17&  0.00&  0.00& -1.378&0.00\\ 
         0.00&  0.33&  0.50&  0.00& -1.378&0.00\\ 
         0.88&  0.12&  0.25&  0.00& 0.626&-2.554\\ 
         0.38&  0.38&  0.25&  0.00& 0.626&-2.554\\ 
         0.12&  0.88&  0.75&  0.00& -0.626&2.554\\ 
         0.62&  0.62&  0.75&  0.00& -0.626&2.554\\ \hline
    \multicolumn{6}{|c|}{+  $\boldsymbol{\tau}_{xy}\mathcal{T}$ }\\
    \hline 
    \end{tabular} 
    \caption{Initial magnetic moments and fractional positions for the Mn atoms with finite magnetic moment in the coplanar structure except for those related by $\boldsymbol{\tau}_{xy}\mathcal{T}$ with  $\boldsymbol{\tau}_{xy}=(1/2,1/2,0)$.\label{sup: 1d}}
\end{table}

Atomic positions remain the same as in the non-coplanar structure and the magnetic moments associated to each site can be found in Tab. \ref{sup: 1d}, where only the positions of the Mn atoms with finite moments are included except for those related by $\boldsymbol{\tau}_{xy}$T with  $\boldsymbol{\tau}_{xy}=(1/2,1/2,0)$. As before, the initial magnetic moments are larger than the resulting ones after the calculation. The rest of the parameters are kept as described above. The DFT result for the total energy is $E_T = -254.44$ eV, that is, a difference of $5$ meV/at with respect to the non-coplanar structure. The band structure can be found in \ref{fig:bands}.

\subsection{Shift photoconductivity spectra of Mn$_{5}$Si$_{3}$}
\begin{figure}
    \centering
    \includegraphics[width=1\linewidth]{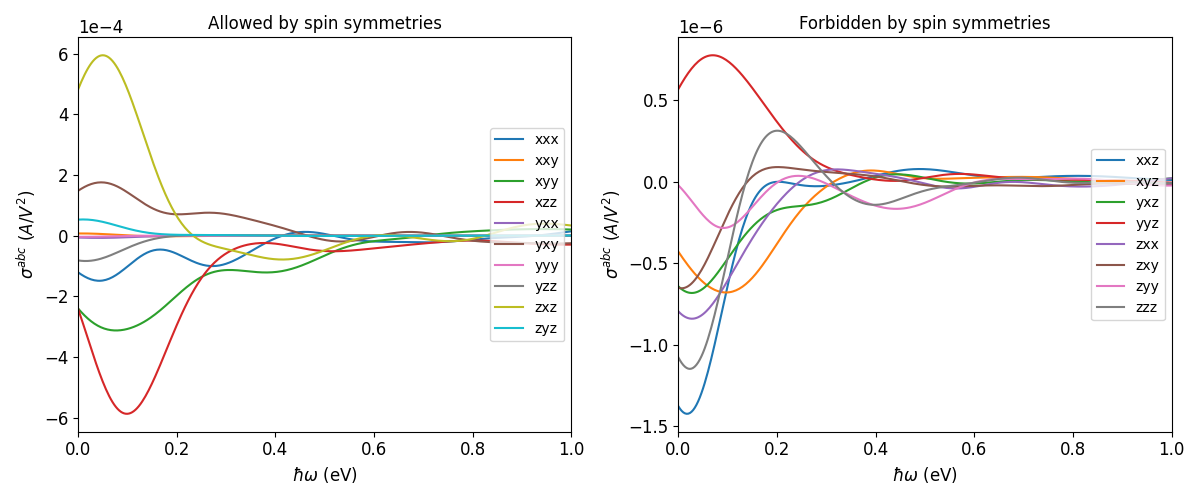}
    \caption{Shift photoconductivity spectrum of the non-coplanar phase of Mn$_{5}$Si$_{3}$ illustrating the effect of spin symmetries. Note the overall
    difference of two orders of magnitude between left and right panels.}
    \label{fig:noncoplanar_sc_sup}
\end{figure}

\paragraph{Non-coplanar.}
Fig. \ref{fig:noncoplanar_sc_sup} displays all 18 components of the shift photoconductivity tensor, separated between those permitted and those forbidden 
(odd instances of $z$ in $abc$)
by spin symmetry considerations.
The figure shows that the forbidden ones are at least two orders of magnitude smaller
than most of the allowed ones. 

\begin{figure}
    \centering
    \includegraphics[width=1\linewidth]{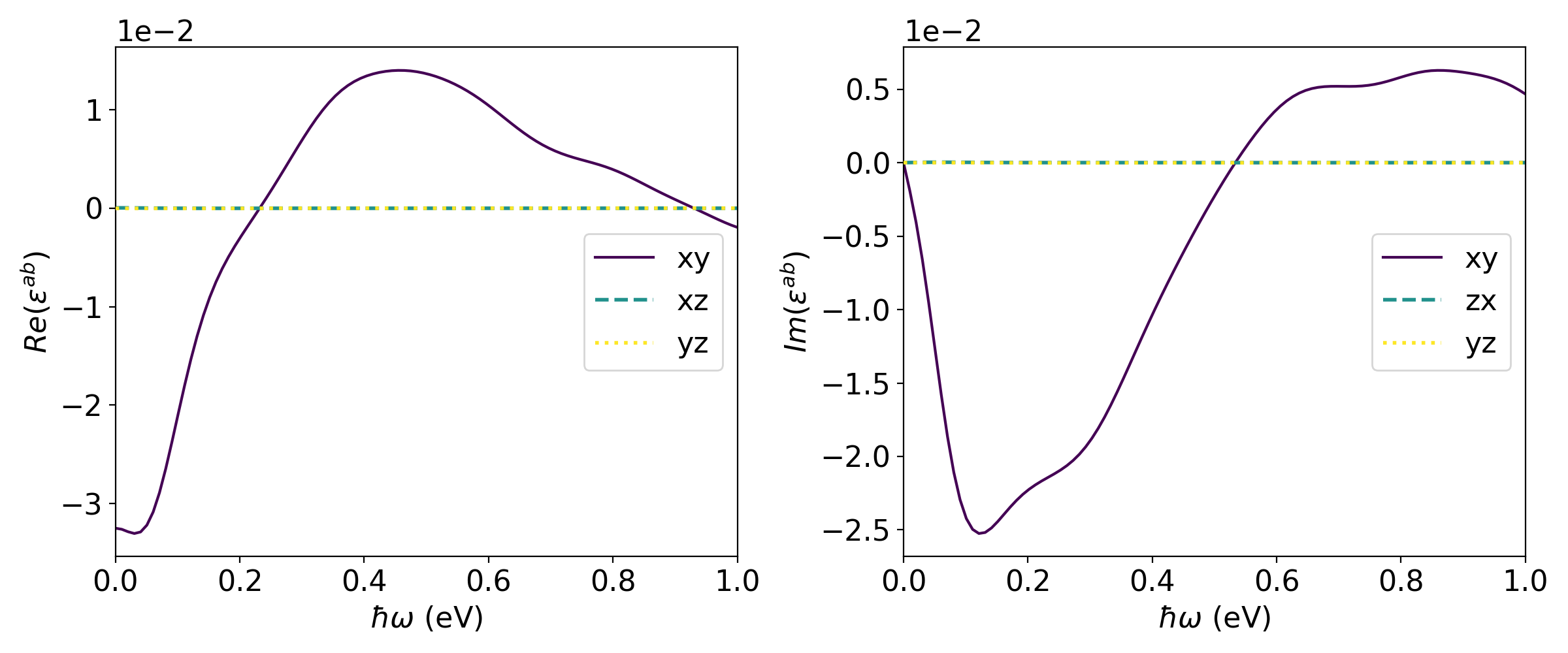}
    \caption{Dielectric tensor of the non-coplanar phase of Mn$_{5}$Si$_{3}$ in units of the vacuum permittivity, illustrating the difference between the spin-symmetry allowed ($xy$) and the forbidden ($zx$ and $yz$) components.}
    \label{fig:dielectric}
\end{figure}

To further illustrate the impact of spin symmetries, Fig. \ref{fig:dielectric} displays the linear optical response tensor, where a distinction is observed between the spin-symmetry allowed off-diagonal component $xy$ and the forbidden components $xz$ and $yz$.

\begin{figure}
    \centering
    \includegraphics[width=0.7\linewidth]{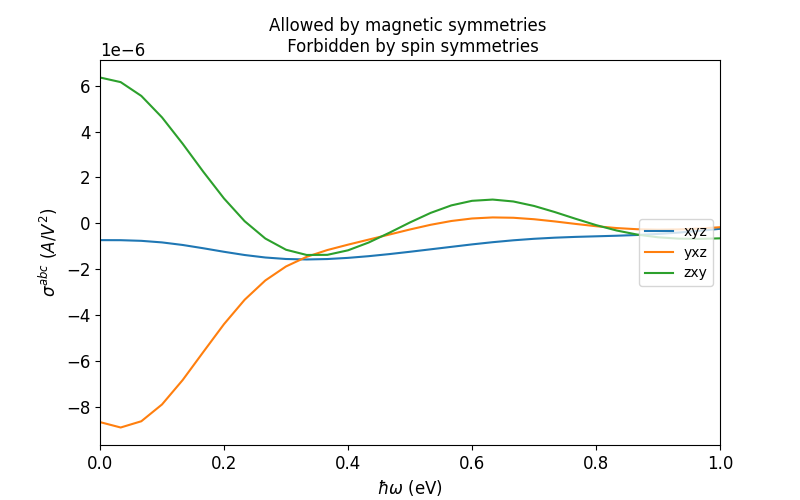}
    \caption{Shift photoconductivity spectrum of the coplanar phase of 
    Mn$_{5}$Si$_{3}$. The overall magnitude is two orders
    smaller as compared to the spin-symmetry allowed components of the non-coplanar configuration in Fig.~\ref{fig:noncoplanar_sc_sup}.}
    \label{fig:coplanar_sc_sup}
\end{figure}

\paragraph{Coplanar.}
In the case of the coplanar structure, spin symmetries predict the vanishing of all tensor components, whereas from magnetic symmetry considerations $\sigma^{xyz}$, $\sigma^{yxz}$ and $\sigma^{zxy}$ are allowed. Fig \ref{fig:coplanar_sc_sup} shows that the values of the photoconductivity  are more than an order of magnitude lower  than most of the allowed components of the non-coplanar structure. 

\subsection{Details on the calculation of the current}


In order to calculate the photocurrent from quadratic optical responses, the expression 
for a sample of thickness $d$ illuminated by  a laser with 
spot size of width $w$
reads~\cite{PhysRevB.101.045104, PhysRevB.108.165418}
\begin{eqnarray}\label{eq:photocurrent}
    J^{a}_{\text{shift}} & = G^{abb}(\omega)\cdot(1-R(\omega))(1- e^{-\alpha^{bb}\cdot d})\cdot w \cdot I^{b},\label{sup:current}
\end{eqnarray}
Above, $I^{b}=c\epsilon_{0} (E^{b})^{2}/2$ is the incident light intensity
for an applied electric field $E^b$,   $R(\omega)$  the reflectivity and
$ \alpha^{bb}(\omega) = \sqrt{2}\frac{\omega}{c}\sqrt{|\epsilon^{bb}|-\text{Re}(\epsilon^{bb})}$.
Finally, 
$
    G^{abb}(\omega)=2\sigma^{abb}(\omega)/c\epsilon_0 \sqrt{\epsilon_r}\cdot\alpha^{bb}(\omega) \label{eq:Glass}
$
stands for the Glass coefficient, which 
quantifies the generation of photocurrents in bulk materials
taking absorption into account~\cite{glass-apl74,tan-cm16}. For simplicity, in our calculations we have set $R(\omega)$ = 0, so the magnitude of the calculated current is expected to be overestimated.
Given our interest in tensor components of the form $\sigma^{abc}$, we provide details on their relationship to the Glass coefficient below.

We  consider the situation described in the main text, namely
current generated along $J^{x}$ under linearly polarized electric field $\mathbf{E}=E_0(0, \cos(\theta), \sin(\theta))$.
For this, we perform a change of coordinates $\mathbf{r}' = \underline{\underline{R}}(-\theta)\cdot\mathbf{r}$ such that the electric field is  polarized purely along the $y'$ direction, that is

\begin{eqnarray}
    \begin{pmatrix}
    0 \\ E_0 \\ 0
    \end{pmatrix}
    = 
    \begin{pmatrix}
    1 & 0 & 0 \\
     0 & \cos(-\theta) &  -\sin(-\theta)  \\
    0  &\sin(-\theta)  &  \cos(-\theta)

    \end{pmatrix}
    \cdot
    \begin{pmatrix}
  0\\ E_0\cos(\theta) \\ E_0\sin(\theta) \\ 
    \end{pmatrix}
\end{eqnarray}

Now, in this reference frame, since there is only intensity $I_b$ in the $y'$ direction and noting that $x'=x$ then
\begin{eqnarray}
    J^{x'}_{\text{shift}} & = G^{x'y'y'}(\omega)\cdot(1-R(\omega))(1- e^{-\alpha_{y'y'}\cdot d})\cdot w \cdot I_{y'} = J^{x}_{shift},\label{sup:current}
\end{eqnarray}
and thus the expression of $G^{abb}$ from the literature can be used except the fact that it is in the rotated frame~\cite{PhysRevB.101.045104, PhysRevB.108.165418}:
\begin{eqnarray}
    G^{xy'y'}=2\sigma^{x'y'y'}(\omega)/c\epsilon_0 \sqrt{\epsilon_r}\cdot\alpha_{y'y'}(\omega).
\end{eqnarray}
The rotated components can be related to the usual ones as
$\sigma^{a'b'c'} = R^{a'a}R^{b'b}R^{c'c} \sigma^{abc}$
and 
$\epsilon^{b'b'} = R^{b'a}R^{b'b} \epsilon^{ab}$.
In our case,
$\epsilon^{y'y'} = \cos^{2}\theta~\epsilon^{yy} + \sin^{2}\theta~\epsilon^{zz}+ \sin^{2}\theta~\epsilon^{yz}$
and
$ \sigma^{xy'y'} = \cos^{2}\theta ~\sigma^{xyy} 
    + \sin^{2}\theta ~\sigma^{xzz} +\sin2\theta ~\sigma^{xyz}$,
leading to
\begin{eqnarray}
        G^{xy'y'}=2 \frac{\cos^{2}\theta ~\sigma^{xyy} 
    + \sin^{2}\theta ~\sigma^{xzz} +\sin2\theta ~\sigma^{xyz}}{c\epsilon_0 \sqrt{\epsilon_r}\cdot\alpha_{y'y'}(\omega)}
\end{eqnarray}
with 
\begin{eqnarray}
\alpha^{y'y'} = \sqrt{2}\frac{\omega}{c}\sqrt{|\cos^{2}\theta~\epsilon^{yy} + \sin^{2}\theta~\epsilon^{zz}+ \sin^{2}\theta~\epsilon^{yz}|-\text{Re}(\cos^{2}\theta~\epsilon^{yy} + \sin^{2}\theta~\epsilon^{zz}+ \sin^{2}\theta~\epsilon^{yz})}
\end{eqnarray}

\subsection{Spin group symmetries}

Here we demonstrate how spin symmetries can impose additional constraints to photoconductivity tensors. 
For that we consider a spin point group operation $g = [R_{s}||R_{r}]$ where $R_{s}$ and $R_{r}$ are proper or improper rotations in spin and real space respectively, that is:
\begin{eqnarray}
            &\mathbf{r} \xrightarrow{g} R_r \mathbf{r}, \\
            &\mathbf{S} \xrightarrow{g} Det(R_{s}) R_{s} \mathbf{S}.
\end{eqnarray}
where $\mathbf{S}$ is the spin operator.
Because $\mathbf{r}$ ($\mathbf{S}$) has only spatial (spin) degrees of freedom is only affected by the real (spin) space operation. Note the appearance $Det(R_{s})$ because of the fact that spin is an axial vector.

\subsubsection{Charge photoconductivity}

In order to study the symmetry properties of the second order charge photoconductivity $\sigma^{abc}$ we need to find how the tensor transforms under a spin operation $g$.
More concretely we will consider the shift photoconductivity tensor analyzed in the main text, which is $\mathcal{T}$ even and symmetric in $b \leftrightarrow c$ and characterized by the equation $j^{a}  = \sigma^{abc} E^{b} E^{c}$. 
Therefore it depends on how both electric field and current transform under a spin operation, which they do in the same way as they are both polar vectors defined in real space.
\begin{eqnarray}
    &\mathbf{E} \xrightarrow{g} R_r \mathbf{E} \\
    &\mathbf{j} \xrightarrow{g} R_r \mathbf{j}    
\end{eqnarray}
The only difference is that the electric field is even and the current is odd under $\mathcal{T}$.

Now we consider two coordinate systems related by the spin operation $g$ such that 
\begin{equation}
    \mathbf{E}^{\prime} = g \mathbf{E} = R_r \mathbf{E}.
\end{equation}
or  $E^{a^{\prime}} = R_{r}^{a^{\prime}a} E^{a}$ component-wise.
Due to the orthonormality of rotation matrices the reverse operation becomes $E^{a}= R_{r}^{a^{\prime}a} E^{a^{\prime }}$.
The equation of interest in both systems read
\begin{eqnarray}
         j^{a}  &= \sigma^{abc} E^{b} E^{c} \\
         j^{a^{\prime}} &= \sigma^{a^{\prime}b^{\prime}c^{\prime}} E^{b^{\prime}} E^{c^{\prime}}. \label{sup:eq1}\\
\end{eqnarray}
Then
\begin{eqnarray}
    j^{a^{\prime}} =R_{r}^{a^{\prime}a} j^{a} = & R_{r}^{a^{\prime}a} \sigma^{abc} E^{b} E^{c} \\
                                      =& R_{r}^{a^{\prime}a} \sigma^{abc} R_{r}^{b^{\prime}b} E^{b^{\prime}} R_{r}^{c^{\prime}c}  E^{c^{\prime}}   \label{sup:eq2}\\
\end{eqnarray}
Comparing Eqs. \ref{sup:eq1} and \ref{sup:eq2} we find 
\begin{eqnarray}
    \sigma^{a^{\prime}b^{\prime}c^{\prime}} = R_{r}^{a^{\prime}a} R_{r}^{b^{\prime}b} R_{r}^{c^{\prime}c} \sigma^{abc} \label{sup:charge}
\end{eqnarray}
Eq.  \ref{sup:charge} coincides with the same known relation for conventional point group operations, except that in this case $R$ does not correspond to a point group operation of the system but to the spatial part of a spin symmetry.
In other words, the magnetic structure does not have to be invariant under $R_{r}$ but under $g=[R_{r}||R_{s}]$, meaning that the set of $R_{r}$ that take part in the spin point group act as an \textit{effective} point group.

\paragraph*{Example.}To examine how Eq. \ref {sup:charge} determines which components are spin-symmetry-allowed we consider  as a concrete example the model from Fig. \ref{fig:model}.
The generators of the spin point group are   $[E || E ]$,  $[\mathcal{T}|| E ]$ and $[E || M_{x} ]$, which only contain the real space operation $M_{x}$ with matrix representation
\begin{equation}
    M_{x} = \begin{pmatrix}
        -1 & 0 \\
        0 & 1
    \end{pmatrix} 
\end{equation}
Invoking Neumann's principle, if a crystal is invariant with respect to certain symmetry elements, any of its physical properties must also be invariant with respect to the same symmetry elements, impliying 
\begin{equation}
        \sigma^{abc} \xrightarrow{g} \sigma^{a'b'c'} =  \sigma^{abc} \label{sup:Neumann}
\end{equation}
Using Eqs. \ref{sup:charge} and \ref{sup:Neumann} we find
\begin{eqnarray}
    \sigma^{xxx} =& \sigma^{x'x'x'} = R_{r}^{x'x} R_{r}^{x'x} R_{r}^{x'x} \sigma^{xxx} = -\sigma^{xxx} \\
    \sigma^{xxy} =& \sigma^{x'x'y'} = R_{r}^{x'x} R_{r}^{x'x} R_{r}^{y'y} \sigma^{xxy} = \sigma^{xxy} \\ 
    \sigma^{xyy} =& \sigma^{x'y'y'} = R_{r}^{x'x} R_{r}^{y'y} R_{r}^{y'y} \sigma^{xyy} = -\sigma^{xyy} \\ 
    \sigma^{yxx} =& \sigma^{y'x'x'} = R_{r}^{y'y} R_{r}^{x'x} R_{r}^{x'x} \sigma^{yxx} = \sigma^{yxx} \\ 
    \sigma^{yxy} =& \sigma^{y'x'y'} = R_{r}^{y'y} R_{r}^{x'x} R_{r}^{y'y} \sigma^{yxy} = -\sigma^{yxy} \\
    \sigma^{yyy} =& \sigma^{y'y'y'} = R_{r}^{y'y} R_{r}^{y'y} R_{r}^{y'y} \sigma^{yyy} = \sigma^{yyy},
\end{eqnarray}
that is, the only allowed components are $\sigma^{xxy}=\sigma^{xyx}$, $\sigma^{yxx}$ and $\sigma^{yyy}$ (see Fig. \ref{fig:sup_model}), contrasting with the prediction from the magnetic symmetry point of view under which all photoconductivity tensor components would be allowed.

\subsubsection{Spin photoconductivity}

We extend now this analysis to spin photocurrents. 
Let's consider the spin shift current $\xi^{s,abc}$, which generates spin current $j^{s,a} = \xi^{s,abc}E^{b}E^{c}$ under circularly polarized light, were $s$ is the spin component. 
This four rank tensor is invariant under $\mathcal{T}$ and antisymmetric under $b \leftrightarrow c$.

The difference with respect to the charge photoconductivity is that $j^{s,a}$ is a spin current which transforms as the product of the spin polarization vector and the current. 
As a result, under a spin group operation g:
\begin{equation}
        j^{s',a'} \xrightarrow{g} Det(R_{s}) R_{s}^{s's} R_{r}^{a'a} j^{s,a}
\end{equation}
Doing the same procedure as above results in 
\begin{equation}
\xi^{s^{\prime},a^{\prime}b^{\prime}c^{\prime}} = Det(R_{s}) R_{s}^{s^{\prime}s} R_{r}^{a^{\prime}a} R_{r}^{b^{\prime}b} R_{r}^{c^{\prime}c} \xi^{s,abc}. \label{sup:spin}
\end{equation}

Equation \ref{sup:spin} reveals a distinct behavior from the magnetic symmetry case; while magnetic symmetries transform all indices uniformly, here we observe different transformations for the spatial and spin indices. Moreover, unlike the charge case, this behavior cannot be described using an effective point group symmetry.

\paragraph*{Example.} We continue the same example from last section (Fig. \ref{fig:model}) but now the spin part of the spin symmetry plays a part.
In this case, the matrix representation of the identity and time reversal operations is $R_{s}=\pm \mathbb{I}$, implying that the only tensor components with the same spin component are related, as $Det(R_{s}) R_{s}^{ij} = \delta_{ij}$.

Then, proceeding in the same manner as above 
\begin{eqnarray}
\xi^{x,xxy} =& \xi^{x',x'x'y'} = Det(R_{s}) R_{s}^{x'x} R_{r}^{x'x} R_{r}^{x'x} R_{r}^{y'y} \xi^{x,xxy} = \xi^{x,xxy} \\ 
\xi^{x,yxy} =& \xi^{x',y'x'y'} =  Det(R_{s}) R_{s}^{x'x} R_{r}^{y'y} R_{r}^{x'x} R_{r}^{y'y} \xi^{x,yxy} = - \xi^{x,yxy}     \\
\xi^{y,xxy} =& \xi^{y',x'x'y'} = Det(R_{s}) R_{s}^{y'y} R_{r}^{x'x} R_{r}^{x'x} R_{r}^{y'y} \xi^{y,xxy} = \xi^{y,xxy} \\ 
\xi^{y,yxy} =& \xi^{y',y'x'y'} =  Det(R_{s}) R_{s}^{y'y} R_{r}^{y'y} R_{r}^{x'x} R_{r}^{y'y} \xi^{y,yxy} = - \xi^{y,yxy}     \\
\xi^{z,xxy} =& \xi^{z',x'x'y'} = Det(R_{s}) R_{s}^{z'z} R_{r}^{x'x} R_{r}^{x'x} R_{r}^{y'y} \xi^{z,xxy} = \xi^{z,xxy} \\ 
\xi^{z,yxy} =& \xi^{z',y'x'y'} =  Det(R_{s}) R_{s}^{z'z} R_{r}^{y'y} R_{r}^{x'x} R_{r}^{y'y} \xi^{z,yxy} = - \xi^{z,yxy} 
\end{eqnarray}
Recalling the tensor is antisymmetric in $b\leftrightarrow c$, the only components that survive are $\xi^{x,xxy}=-\xi^{x,xyx}$, $\xi^{y,xxy}=-\xi^{y,xyx}$ and $\xi^{z,xxy}=-\xi^{z,xyx}$ (see Fig. \ref{fig:model} e)

This situation is fundamentally different when viewed from the perspective of magnetic symmetry. Here, not only is the magnetic result entirely distinct (lacking any symmetry, which implies that every component should be allowed), but even if the magnetic point group were identical to what we refer to as the \textit{effective} point group, its behavior would still differ. For example, in such a case, $\xi^{y,xxy}$ would be zero.

\end{document}